\documentclass[a4paper,alpha-refs]{aejstyles}

\journal{aej}

\usepackage{graphicx}
\usepackage{siunitx}
\usepackage{multirow}
\usepackage{textcomp}
\usepackage[left]{lineno}

\title{How Do Observational Astronomers Learn to Inspect Imaging Data?}

\papercat{Research Article}

\author[1,\authfn{1}]{H. K. Walsh}
\author[1]{C. J. Fluke}
\author[1]{S. Webb}
\author[2]{L. Wise}

\affil[1]{Centre for Astrophysics \& Supercomputing, 
Swinburne University of Technology, Hawthorn, 3122, Australia}
\affil[2]{Department of Psychological Science, 
Swinburne University of Technology, Hawthorn, 3122, Australia}

\jvolume{00}
\jnumber{0}
\jyear{2024}

\begin{document}

\begin{frontmatter}
\maketitle
\begin{abstract}
Astronomy is entering an unprecedented era of data collection. Upcoming large surveys will gather more data than ever before, generated at rates requiring real-time decision making. Looking ahead, it is inevitable that astronomers will need to rely more heavily on automated processes. Indeed, some instances have already arisen wherein the majority of the inspection process is automated. Visual discovery, performed traditionally by humans, is one key area where automation is now being integrated rapidly. Visual discovery comprises two aspects: (1) visual inspection, the skill associated with examining an image to identify areas or objects of interest; and (2) visual interpretation, the knowledge associated with the classification of the objects or features. Both skills and knowledge are vital for humans to perform visual discovery, however, there appears to have been limited investigation into how the skill of visual inspection in astronomy is acquired. In this work, we address this issue by setting out to identify the landscape within which observational astronomers develop the skills to perform visual inspection. We report on a survey of 70 professional observational astronomers, at various career stages and from different geographical regions. We found that between 63\%and 73\% of the astronomers surveyed had received formal and/or informal training in visual inspection of images, although formal training (21\%) was less common than informal training (60\%). Surprisingly, out of the 37\% who did not recall having received training in visual inspection, 29\% (20 astronomers) indicated that they provided training to others.  This suggests the emergence of ``expertise without precedent'' where a first expert in the field provides a new way of achieving a task. These results, paired with a set of three pilot interviews, present a touchstone against which the training of future observational astronomers can be compared.
\end{abstract}


\end{frontmatter}


\section{Introduction}
In its earliest form, observational astronomy was dependent on what an astronomer could see, record, classify or share with others (e.g. through oral traditions, story-telling, and performance). Advancements in technology in observational astronomy- from telescopes, spectroscopy and photography to charged coupled devices, radio interferometers and digital archives- have focused on improving the astronomer's ability to perform one or more of these tasks.

The 21st century marked a subtle shift. While the creation of new technologies is still a significant driver of advancement in astronomy we are simultaneously running into an altogether different problem: the limitations of human ability. The new era of so called big data or intensive astronomy (see for example, \citet{Ball10} or \citet{berriman13}) is changing the way astronomers engage in the research process. This shift can perhaps most prominently be seen in the way astronomers undertake the process of visual discovery when looking at astronomical images.

\subsection{Defining Visual Discovery}
\label{defining visual discovery}

As an observational science, astronomy is dependent on the collection and analysis of images or image-like data products.  This includes optical imaging, using wavelength ranges that overlap with the human visual system, and data collection in other bands of the electromagnetic spectrum, where data values are rendered as images that can be viewed by an astronomer.

Visual discovery is an integral competent of research workflows where astronomers generate new knowledge from astronomical images.  In this work, \textbf{visual discovery} is defined as a process wherein an image is inspected and objects, areas, or features of interest are identified and then interpreted through activities such as evaluation, classification, or data analysis.

From this definition, visual discovery comprises two separate  yet interlinked cognitive activities that the (human) astronomer must undertake: 

\begin{itemize}
\item \textbf{Visual inspection} is the act of examining an image to identify areas or objects of interest. Performing visual inspection is largely dependent on the {\em skills} an observer has at scanning over or navigating their way within an image; and
\item \textbf{Visual interpretation} involves the classification of the objects or features that have been found. The quality or effectiveness of visual interpretation is largely dependent on the {\em knowledge} an astronomer has accumulated over their career. 
\end{itemize}

This differentiation drawn between visual inspection and visual interpretation is broadly consistent with psychological models of biologically primary and secondary knowledge \citep{Geary08}. Moreover this distinction appears in other disciplines, for example medical diagnostic imaging, where there is an acknowledged difference between visual inspection and interpretation (see for example \citet{vanderGijp2014}). The intersection of training in astronomy and other visual-dominated fields is an important discussion that we will return to in Section \ref{summary and discussion}.

\subsection{Types of Training}
The definitions of training/learning models presented here are informed by reviews such as \citet{Manuti15} and \citet{colley02}. The duality of research as both vocation and education requires certain criteria to be excluded from the provided definitions (e.g. education at an institution does not suggest a degree of formality in this context). Additionally, a significant portion of research occurs at, or is performed in association with research institutions. Therefore, non-formal training has been excluded as a mode of training.

In this work we categorise training in two broad modes: (1) Formal training is highly structured and has well-defined outcomes (e.g. follows a well-defined syllabus or curriculum with specific learning objectives) that can be assessed. Just-in-case education and the one-to-many delivery mode of lectures or tutorial classes are predominantly formal approaches to training. (2) Informal training is largely unstructured, and training activities may be specific to the trainee's needs. Mentor-mentee, such as occurs between a PhD supervisor and a postgraduate student, and self guided learning are predominantly informal approaches to training.

\subsection{Automation in Visual Discovery}
\label{automation}

Automation in astronomy includes systems such as machine learning, or other implementations of artificial machine intelligence, with an emphasis on data processing pipelines [e.g. \citet{davies13}, \citet{crispred15}, \citet{bosch18}, \citet{masci2019}, \citet{weilbacher20}, and \citet{leroy21}] and source finding systems [e.g. \citet{bertin96}, \citet{Makov05}, \citet{hales12}, \citet{riggi16}, \citet{robotham18}, and \citet{serra15}]. One of the primary aims of automation is to respond to the challenges of growth in both the volume and velocity of data in astronomy.

Automation as a component of the visual discovery process is not a new concept. While we might consider the Sloan Digital Sky Survey \citep{York2000} or the emergence of the `Virtual Observatory' \citet{brunner98} to be early drivers of digital automation, arguably the history of automation in observational astronomy goes as far back to, if not further than, the work of the  Harvard `Human Computers' \citep{Kelly18}. 

Automation within observational astronomy typically addresses the data volume problem within visual discovery. For example, SExtractor \citet{bertin96}, has an almost three decade long history of performing source detection. SExtractor was not the first source detection software but it is a significantly long lasting one. From the definition of visual discovery above, the 1996 version of SExtractor could be described as using: (1) thresholding to perform visual inspection; and (2) a neural network to perform visual interpretation, distinguishing between stars and galaxies. While this approach is effective for building catalogues that support statistical discoveries, to an extent it diminishes the opportunities for unplanned discoveries which may not fit into our models \citep{Norris17}.

As an illustrated example of how automation is changing visual discovery, consider the role of automation in optical transient astronomy through programs such as the Palomar Digital Sky Survey \citep{Djorgovski1998}, the Palomar Transient Factory \citep{Law2009}, the Zwicky Transient Facility \citep{Bellm2014}, the Dark Energy Survey \citep{Abbott2016}, the All-Sky Automated Survey for SuperNovae \citep{Kochanek2017}, and the Deeper Wider Faster Program \citep{andreoniandcooke}.

In optical transient astronomy, astronomers are searching for sources that change or evolve over time. Optical transient astronomy often involves looking at:
\begin{enumerate}
    \item  Difference imaging of the source location; and
    \item Light curves of the sources as they change in magnitude over time.
\end{enumerate}

Optical transient programs have utilized custom user interfaces, databases and even room-scale setups to allow rapid visual inspection and interpretation of this data \citep[e.g.][]{Law2009,Meade17,Kochanek2017,masci2019}. 
    
Visual discovery is a vital part of such surveys: automation facilitates the isolation of evolving targets both at a scale and within a time-frame that would otherwise not be possible. While many of these programs utilise machine learning algorithms to support astronomers, for now, the ultimate decision to classify a source as interesting is often left to a human.

Further, as we enter the era of the Vera Rubin Legacy Survey of Space and Time \citep[LSST,][]{LSST2009}, with an expected candidate rate of 10 million per night, automated data brokers are being designed to streamline visual discovery for astronomers \citep{Narayan_2018}. An inevitable outcome of engaging in larger, faster, real time surveys is the reliance on more sophisticated automated systems. 

What makes optical transient astronomy an instructive case is that it is the data velocity, rather than the data volume, that presents the main challenge. As such the physiological limitations of researchers -the inability to keep up with the data rate- is the key driving force for automation. Additionally, it poses important questions for next-generation observational astronomy. As other observational fields approach the rate of data generation already present in transient astronomy, will similar levels of automation become the norm? And what does this mean for the skill of visual inspection in observational astronomy?

\subsection{Cyber-Human Discovery Systems}
One proposed solution to the balancing of human and automated systems is the cyber-human discovery system discussed in \citet{Fluke20b}. Here, the authors suggest strategies to support a strong synergistic relationship between astronomers and automated systems in order to enhance the performance and outcomes of both. Four key areas where automation could enhance the quality of the researcher's performance in visual inspection were proposed: (1) talent identification; (2) identifying and improving visual search strategies; (3) adaptive user interfaces; and (4) provision of just-in-time coaching. Prior to implementing any of these approaches, it is important to understand how astronomers currently attain  their skills in visual inspection.  

\subsection{Development of Knowledge and Skills}
\label{Knowledge}
In observational astronomy there have been very few studies that investigate how the professional cohort performs visual discovery with images. An eye-tracking study by\citet{Arita11} suggests differences in the fixations of expert and novice astronomers when inspecting images. \citet{meade14} compared image search performance (speed and accuracy) between cohorts of astronomers and non-astronomers while using either a standard desktop display or a very high resolution tiled display wall. They found that astronomers had better performance overall but with significant individual differences. \citet{Rojas23} studied interpretation performance in gravitational lens observation, based on the proportion of correctly-identified gravitational lenses. With a cohort comprised of astronomers at different career stages, as well as citizen scientists\footnote{A citizen scientist is a member of the general public who is assisting in a collaborative research project.}, they found no significant difference in the ability to perform classifications based on experience or career stage.

Skills in visual inspection and knowledge in visual interpretation play a symbiotic and cyclical role in the process of visual discovery. We might expect that two astronomers with equivalent knowledge will interpret a feature in a similar way (e.g. both astronomers agree that a feature with a strong central concentration and prominent spiral arms is a spiral galaxy). We would also expect that as an astronomer advances in their career they accrue more knowledge and will therefore be able to interpret more features. The ability to perform visual interpretation is easily tested, either an individual can or cannot, identify an object based on the presents of given features. For example, an individual may be able to visually identify an area with higher density of stars (spiral arms) but without sufficient knowledge of astronomy they cannot interpret that information further. \citet{Eriksson17}  refer to this concept in terms of disciplinary discernment, where the ability to discern an object is based on knowledge. In comparison, despite the recognition of the importance of visual inspection in observational astronomy, it is still unclear how skills in visual inspection are developed or what effect these skills have on the discovery process.

\subsection{Overview}
Visual inspection is a fundamental part of observational astronomy. Despite its perceived importance, the actual methods by which skills in visual inspection are currently developed are unclear. In the past visual inspection was a skill that could be ``learned by doing''. As the degree of automation in observational astronomy is increasing, opportunities to ``learn by doing'' are decreasing. To move towards a future where automation is used to enhance rather than replace the researcher requires careful consideration of the role of the researcher. 

In this section we have evaluated the role of visual discovery within observational astronomy, considered the changing landscape of automation in astronomy and discussed the potential effects future automation could have on visual discovery. What remains unclear is the method through which astronomers develop skills in visual inspection.

The purpose of this paper is therefore to present data on the training experience of observational astronomers. To gather this data, a survey was conducted between June 2021 and May 2022, the results for which are outlined in Sections \ref{demographics} and \ref{training in visual inspection}. Additionally the results of three pilot interviews are presented in Section  \ref{interviews on iddividual experience of training}. Finally, in Sections \ref{summary and discussion} and \ref{final thoughts} a reflection on the collected results is presented.

The lack of clarity in the development methods of skills in visual inspection paired with the increasing utilisation of automation in visual inspection present significant questions for educators: To what extent do educators need to provide training in visual inspection? And, in a landscape where opportunities to perform visual inspection are diminishing, how do educators provide training in visual inspection?

The answers to these questions will perhaps only be clear once they evolve over time. However, through this work we aim to promote discourse on visual inspections historic role in astronomy as well as evaluating what its role is as astronomy continues to become a more data intensive area of research. Importantly we advocate for documentation of the methods through which both formal and informal  training is provided in order to better isolate high quality training practices.

\section{Investigating Training in the Skill of Visual Inspection} \label{Investigating Visual inspection in Modern Astronomy}
In order to better understand how astronomers develop their skills in visual inspection of images, an online survey was developed with four focus areas in mind. These were:

\begin{enumerate}
    \item Through which methods is training in visual inspection delivered? Is it provided formally through methods such as structured classes or lectures or informally through methods such as the guidance of a mentor? This is discussed in Section \ref{formal training}.
    \item To what extent do astronomers receive training in visual inspection? Is training only provided to researchers in the early stages of their careers? Or do senior astronomers engage in training to perform skill maintenance? This is discussed in Section \ref{frequency of training}.
    \item What is the perceived value of training astronomers receive in visual inspection? Is there a difference in value depending on the formality of training? Does training remain valuable irrespective of the time since it was delivered? This is discussed in Section \ref{training relevance}.
    \item Who provides astronomers with training in visual inspection? Is training only provided by senior researchers? Are trainers specialists who only provide training or is it a small portion of their workload? This is discussed in Section \ref{the provision of training in visual inspection}.  
\end{enumerate}

\subsection{Survey Questions} 
The survey consisted of 13 questions divided into two sets as shown in Table \ref{tab:questions}. The first set included 5 demographic questions that were used to identify key groups of respondents. In addition, demographic information allowed for the responses regarding training to be sorted by key factors such as career stage. The second set included 8 questions that directly related to training in visual inspection of images. Some minor edits have been made to the formatting of the questions here as compared to the way in which they were presented in the survey (removal of excess capital letters etc.).
\begin{figure}[bt!]

    \centering
    \includegraphics[width = \linewidth]{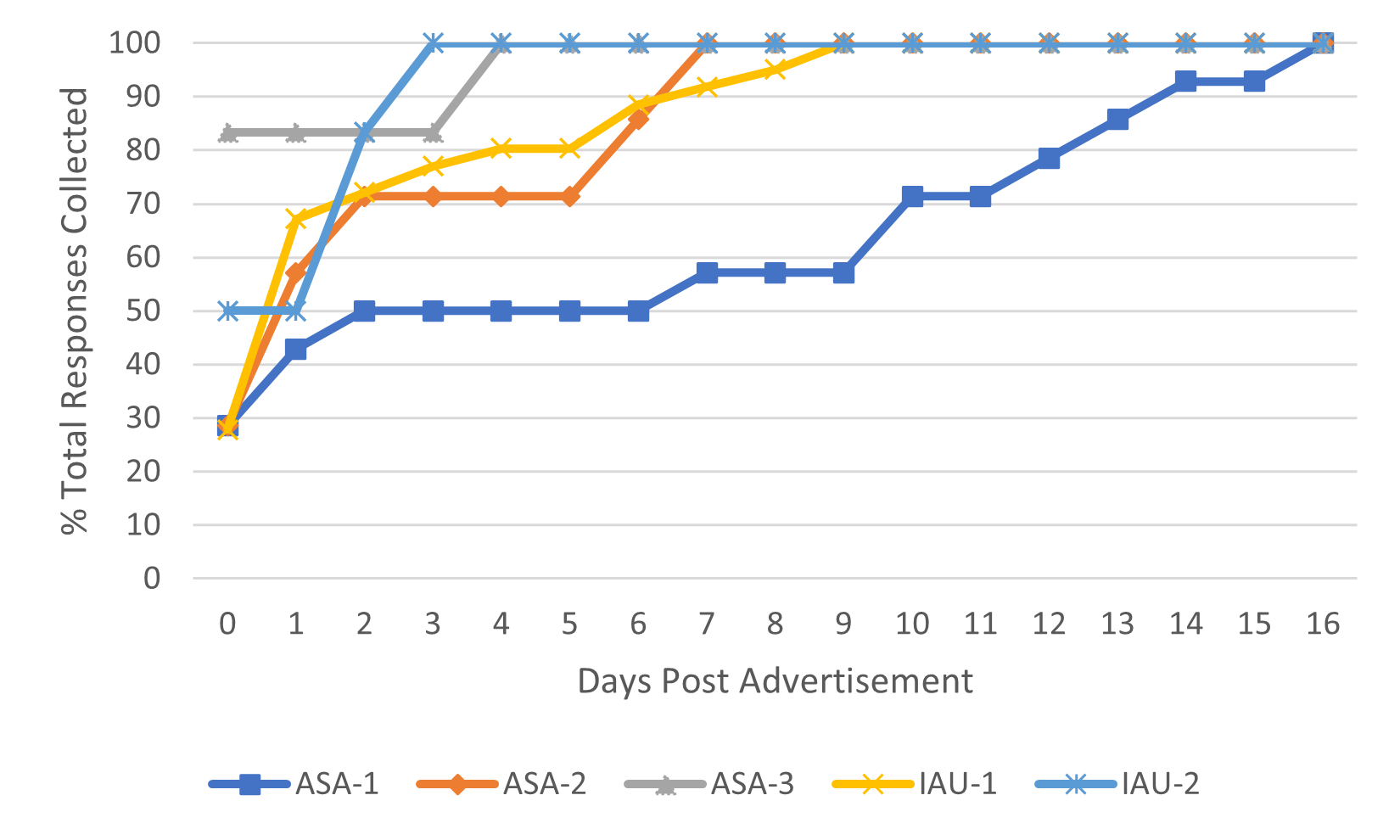}
    \caption{The percentage of the total number of responses collected from each advertising session compared to the number of days post advertising. It is notable that in no case were additional responses attributed to an advertising session after 16 days had passed. The total number of responses received from each advertising session was: ASA-1 = 14 responses, 
    ASA-2 =  7 responses, ASA-3 = 6 responses, IAU-1 = 61 responses and IAU-2 = 6 responses.}
    \label{response figures}

\end{figure}
\begin{table*}[bt!]
    \caption{List of questions presented to respondents in ``Understanding Visual Discovery in Data-Intensive Disciplines''. Questions are listed with reference to sections, tables and figures where they are discussed.}
    \centering
    \begin{tabular}{ccp{0.65\linewidth}p{0.05\linewidth}p{0.06\linewidth}p{0.04\linewidth}}
     ~ & ~ & ~ & Section & Table & Figure \\
       ~   & 1 & Please select the geographical region where you are currently (or were most recently) studying or working. & \ref{Geographic Location} & \ref{tab:geo} \\
       ~ & 2 & What is your current age? &  \ref{career stages} & \ref{tab:age 2 cs}   \\
        ~  & 3 & Which one of the following options best describes your current career stage? & \ref{career stages} & \ref{tab:age 2 cs} \\
        ~ & 4 & Which one of the following options best describes your current or planned area of research? & \ref{area of research} & \ref{tab:obs} \\
        \multirow{-5}{*}{\rotatebox[origin=c]{90}{Demographics}} & 5 & Which one of the following options is the best description of the data set that you are using in your current (or most recently completed) research work?  & \ref{research scale} & \ref{tab:percents}\\ \hline 
        ~ & 6 & Have you ever received formal training in visual inspection of Astronomical images? & \ref{formal training} & \ref{tab:Formal} \& \ref{tab:formalvinformal} & \ref{circles}\\
        ~& 7 &  Have you ever received informal training in visual inspection of Astronomical images? & \ref{formal training} & \ref{tab:informal} \& \ref{tab:formalvinformal} & \ref{circles} \\
        ~ & 8 & When did you last receive formal training in visual inspection? & \ref{frequency of training} & \ref{tab:Timesformal}\\
        ~& 9 &When did you last receive informal training in visual inspection? & \ref{frequency of training} & \ref{tab:Timesinformal}\\
        ~ & 10 & How relevant to your current work is the formal or informal training you have received? & \ref{training relevance} & \ref{tab:relevance} & \ref{training relevance bars}\\
        ~ & 11 & Consider your most recent experience receiving informal training in visual inspection of astronomical images. From the following list, please select the most accurate description of the role of the person who provided that informal training. & \ref{the provision of training in visual inspection} & \ref{tab:who are trainers}\\
        
        ~ & 12 & To what extent do your current (or most recently completed) duties require you to deliver formal training in visual inspection activities? & \ref{the provision of training in visual inspection} &  \ref{tab:untrainedformaltrainers} \& \ref{tab:supply formal training all} & \ref{Supply formal training multi} \\
        
        \multirow{-8}{*}{\rotatebox[origin=c]{90}{Training Experience}} & 13 & To what extent do your current (or most recently completed) duties require you to deliver informal training in visual inspection activities? & \ref{the provision of training in visual inspection} & \ref{tab:untrainedinformaltrainers} \& \ref{tab:supply informal training all} & \ref{Supply formal training multi}  \\
    \end{tabular}

    \label{tab:questions}
\end{table*}

\subsection{Participants}
\label{Distribution and Participation}
Responses to the survey were collected through the Qualtrics online survey platform\footnote{\url{https://www.qualtrics.com/}} from 17 June 2021 to 20 May 2022. 

Initially, this survey was only advertised to the Astronomical Society of Australia's (ASA) membership, which occurred via email on 17 June 2021 and 26 July 2021. After 4 months, the survey received 27 responses from the ASA as can be seen in Figure \ref{response figures}. An additional final advertisement was distributed to the ASA's membership on 3 April 2022, this obtained 6 new responses. To increase the exposure of this project to an international audience, additional advertisements were distributed via email to the International Astronomy Union's (IAU) membership on 3 May 2022 and 17 May 2022.

As can be seen in Figure \ref{response figures}, responses tended to arrive on the day of advertisement and the day after advertising. Within these first 2 days, over 60\% of all responses were collected; after the first week, 84\% of the total responses were collected. Additionally, response numbers indicated that subsequent advertisements to the same cohort generated significantly lower engagement than the original advertisement.

We note the version of the survey advertised to the ASA membership in 2021 contained additional questions, which are discussed in a companion paper on the use of advanced image displays in astronomy [Fluke et al. (submitted)]. Moreover, a request for survey respondents to participate in interviews with the research team was removed when the survey was advertised in 2022 (See Section \ref{interviews on iddividual experience of training}).

\subsubsection{Professional Cohorts}

The ASA is a national professional astronomy organisation that actively encourages postgraduate students to become members. At the time the survey was first advertised (17 June 2021), the ASA had a membership of approximately 700 individuals. From the first round of advertising that was provided to the ASA, the survey received responses from 2\% of its members. In total, from all the advertising provided to the ASA, approximately 3.9\% of members responded to this survey. 

The IAU is the largest global professional astronomy organisation. IAU membership is open to individuals with a PhD or equivalent. At the time the survey was first advertised (3 May 2022), the IAU had a membership of over 12000 individuals. Due to its size, the IAU supports its membership through divisions within key research areas. Two divisions were selected whose research interests correlated with the target audience of this survey. The first group was Division B: ``Facilities, Technologies and Data Science'', with \textasciitilde 4250 members. The second group was Division J: ``Galaxies and Cosmology'', with \textasciitilde 3950 members. IAU members can join multiple divisions and therefore the actual number of potential respondents that were contacted is not known. On the basis that the true number of members contacted must be between \textasciitilde 4250 and \textasciitilde 8200, the IAU had an estimated response rate between 1.5\% and 0.75\%.

\subsubsection{Exclusions}
By the closure data of the survey, 94 responses were obtained comprising 27 collected through advertising to the  ASA and 67 collected through advertising to the IAU. Of these,  12 responses were excluded due to one of two reasons. Firstly, if respondents answered fewer than 50\% of the survey questions then their responses to all questions were excluded, this accounted for 9 exclusions. Secondly, 3 respondents were part of cohorts whose research background or current research was not within the scope of responses that were of relevance for this paper (i.e. astronomers). These exclusions result in a sample size of 82 astronomers. As we explain in Section \ref{area of research} we further reduced our sample for full analysis to the 70 respondents who indicated that they were observational astronomers.

\section{Demographics}
\label{demographics}
In this section, the three key pieces of demographic information that were collected are outlined. These demographics are used to group respondents with similar characteristics, namely: geographic region (from Question 1), career stage (from Question 3) and research area (from Question 4). In addition, two other pieces of demographic information were collected. These demographics were not used for response grouping, namely: age (from Question 2) and scale of research activity (from Question 5).
\subsection{Geographic Region}
\label{Geographic Location}
One aim of advertising through the IAU was to engage with astronomers from differing research backgrounds. The hope was to identify if there were region-specific differences in training. As seen in Table \ref{tab:geo} by extending the survey from the ASA to the IAU a broader sample of international respondents was obtained. Despite capturing a comparable number of responses from Australia and New Zealand, Europe and North America, no clear evidence of region-specific differences in responses was found. consequently, no further analysis is presented relating to geographic region.

\begin{table}
    \caption{\textit{Question 1:  Please select the geographical region where you are currently (or were most recently) studying or working.} While the option was provided no responses were made by astronomers from Africa, the Middle East, or Oceania outside of Australia and New Zealand.}
    \centering
    \begin{tabular}{|l|c|c|c|}
    \hline
        ~ & ASA & IAU & Total \\ \hline
        Australia/New Zealand & 19 & 6 & 25 \\ \hline
        Europe & 1 & 25 & 26 \\ \hline
        North America & 0 & 23 & 23 \\ \hline
        Other & 0 & 8 & 8\\ \hline
        Total & 20 & 62 & 82 \\ \hline
    \end{tabular}

    \label{tab:geo}
\end{table}
\subsection{Career Stage} \label{career stages}
When designing the survey questions, it was hypothesised that the career stage of an individual may be a factor in their participation in on-going training activities. Rather than asking the subjective question: ``What stage of your career do you think you are in?'', a question regarding the years since the award of PhD was selected. While this is not a perfect analogue for expertise in astronomy, as it does not allow for recognition of career breaks, it presents a reasonable metric for considering currency of training. It was also expected that time post-award of PhD would be correlated with participant age. As seen in Table \ref{tab:age 2 cs}, this assumption was accurate. Although not explored here it could also allow for the consideration of career trajectories other than the common ``undergraduate student to researcher'' pathway.

For the purpose of this paper, the following four categories are used to group individuals at similar career stages, noting that ECR combines two subcategories (0-5 years and 5-10 years) that were presented separately to survey participants:
\begin{itemize}
    \item A student who is undertaking postgraduate studies (PGS);
    \item An early career researcher who is 10 or fewer years post award of PhD (ECR);
    \item A mid-career researcher who is between 10 and 20 years post award of PhD (MCR); and 
    \item A senior career researcher who is more than 20 years post award of PhD (SCR).
\end{itemize}

\begin{table}
    \caption{\textit{Question 2: What is your current age?} and \textit{Question 3: Which one of the following options best describes your current career stage?} In this table we use: PGS = post graduate student; ECR = early-career researcher; MCR = mid-career researcher; and SCR = senior career researcher. These career stages are described in Section \ref{career stages}.}
    \centering
    \begin{tabular}{|l|c|c|c|c|c|c|}
    \hline
         ~ &  18-24 & 25-34 & 35-44 & 45-54 & $>$55 & Total \\ \hline
        PGS  & 3 & 11 & 1 & ~ & ~ & 15 \\ \hline
        ECR & ~ & 7 & 11 & ~ & ~ & 18 \\ \hline
        MCR  & ~ & ~ & 13 & 6 & ~ & 19 \\ \hline
        SCR  & ~ & ~ & ~ & 11 & 19 & 30 \\ \hline
        Total & 3&18&25&17&19& 82\\ \hline
    \end{tabular}

    \label{tab:age 2 cs}
\end{table}

\subsection{Research Sub-Disciplines}
\label{area of research}
Almost all astronomers will perform visual discovery during their research activities, however, visual discovery is particularly important for observational astronomers. As such, a key piece of demographic information that was collected was the primary research area of respondents, reported in Table \ref{tab:obs}. Options provided in the survey included 5 observational areas of astronomy: Optical/Infrared, Radio, Gravitational wave/Multi-messenger, Multi-wavelength or Other. Due to the low number of respondents (2), researchers in the field of Gravitational Wave/ Multi-messenger astronomy are reported in the Other category. In addition, there were 5 non-observational areas of astronomy that could be chosen: Astronomical Instrumentation, Computational Astrophysics, Technical Software Development/Applied Computing, Theoretical Astrophysics or Other.

\begin{table}[]
    \caption{\textit{Question 4: Which one of the following options best describes your current or planned area of research?} This table includes all 82 responses.}
    \centering
    \begin{tabular}{|l|p{0.125\textwidth}|c|c|c|c|c|}
    \hline
        ~&~ &PGS & ECR & MCR & SCR &  Total \\ \hline
        ~ & Optical/Infrared & 6 & 9 & 7 & 15 & 37 \\ \cline{2-7}
        ~ & Multi-wavelength & 2 & 4 & 2 & 6 & 14 \\ \cline{2-7}
        ~ &Radio & 2 & 3 & 6 & 3 & 14 \\ \cline{2-7}
        ~ &Other & 1 & 1 & 1 & 2 & 5 \\ \cline{2-7}
       \multirow{-5}{*}{\rotatebox[origin=c]{90}{Observational}} & Observational Total  & 11 & 17 & 16 & 26 & 70 \\ \hline
        ~&Computational   \newline Astrophysics & 1 & 0 & 3 & 0 & 4 \\ \cline{2-7}
        ~&Astronomical  \newline Instrumentation & 0 & 0 & 0 & 1 & 1 \\ \cline{2-7}
        ~&Technical \newline Software \newline Development & 0& 1 & 0 & 0 & 1 \\ \cline{2-7}
        ~&Theoretical \newline Astrophysics & 1 & 0 & 0 & 0 & 1 \\ \cline{2-7}
        ~&Other & 2 & 0 & 0 & 3 & 5 \\ \cline{2-7}
        \multirow{-9}{*}{\rotatebox[origin=c]{90}{Non- Observational}}&Non-Observational Total & 4 & 1 & 3 & 4 & 12 \\ \hline

    \end{tabular}

    \label{tab:obs}
\end{table}

As we expect that astronomers working in observational fields will have the greatest requirements for visual inspection, we further reduce our sample to the 70 respondents whose primary research area is observational. Our analysis of the responses to Question 6-13 now use this sample.

\subsection{Data Sets and Research Scale}
\label{research scale}
A contributing factor in a researcher's ability to undertake visual inspection is the scale of the data set under consideration. In this survey, the scale of data has been measured in terms of ``objects'' in observational research and ``particles, cells or equivalent'' for those utilising numerical simulations as the source of data. A limit for manual visual inspection cannot be presented without knowing factors like the rate at which data must be viewed (viz. data velocity) be viewed and the number of individuals involved in the inspection process, however, observational studies with  $>10^{4}$ objects are highly likely to utilise some form of automation.

Table \ref{tab:percents}  shows that larger-scale projects were more common amongst the respondents. The largest response group was large-scale observational surveys which accounted for 41\% of all responses. This was followed by medium-scale observational survey (24\% of responses) and then small-scale observational survey (23\% of responses). However, it is notable that while few, there are still astronomers who work on single-object studies. Interestingly, 6 non-observational respondents indicated that they worked with observational surveys, similarly, 2 observational respondents indicated working with numerical simulations.

\begin{table}[bt!]
    \caption{\textit{Question 5: Which one of the following options is the best description of the data set that you are using in your current (or most recently completed) research work?} Responses that are observational in nature are reported by ``objects'' whereas numerical responses are by ``particles, cells or equivalent''. This table includes all 82 responses. Responses are separated for observational (Obs) and non-Observational (non-Obs) astronomers.}
    \centering

    \begin{tabular}{|l|l|c|c|c|}
    \hline
        ~ & ~& Scale & Obs & non-Obs \\ \hline
        ~ &  Single object& $1$ & 2&1  \\[0.2cm] \cline{2-5}
                                     &Small-scale & $1 - 10^{2}$  & 18 &1 \\[0.2cm] \cline{2-5}
                                    &Medium-scale & $10^{2} - 10^{4}$ & 18&2\\[0.2cm] \cline{2-5}
        \multirow{-5.5}{*}{\rotatebox{90}{\begin{tabular}{c}Observational \\ Surveys\end{tabular}}}  &Large-scale &  $>10^{4}$ & 30&4 \\[0.2cm] \hline
        ~ & Small-scale &  $<10^{5}$ & 0&1 \\[0.2cm] \cline{2-5}
        ~&Medium-scale & $10^{5} - 10^{9}$ & 1&1 \\[0.2cm] \cline{2-5}
        \multirow{-4.5}*{\rotatebox{90}{\begin{tabular}{c} Numerical \\Simulations\end{tabular}}}&Large-scale &   $>10^{9}$ &1 & 2\\[0.2cm] \hline
        Total &~ & ~& 70&12 \\[0.2cm] \hline

    \end{tabular}

    \label{tab:percents}
\end{table}
\subsection{Summary}
The demographic information that was collected shows that a range of astronomers engaged with this survey. These demographics help establish two factors that are important in the investigation of training, namely: (1) the opportunities for training in visual inspection and; (2) the contexts in which training in visual inspection is applied.

In terms of opportunities for training, Table \ref{tab:geo} shows that a significant number of responses have been obtained from astronomers in regions with 3 to 4 year PhD programs such as Australia and New Zealand, as well as regions with 4- to 6-year PhD programs such as North America. Longer PhD programs could present greater opportunities for formal training programs to be provided, though, as stated in section \ref{Geographic Location} no region-specific differences were found in the responses to this survey. 

The results in Table \ref{tab:age 2 cs} show a relatively even split of PGS, ECR, MCR and SCRs. While more senior researchers may have had a long time to engage with training opportunities, the types of training available to PGS are significantly different.

The aim of Questions 1, 2 and 3 was to sample from a reasonable number of research backgrounds. The responses that have been obtained suggest that this has been achieved.

In terms of training context, Table \ref{tab:obs} show that the majority of respondents worked in observational fields of astronomy. The two divisions that were targeted through the IAU were specifically chosen because their research areas were highly related  to observation. While the methods through which non-observational astronomers are trained in visual inspection are also of interest, it is not the focus of this paper. In regard to observational astronomers: while the key components of visual inspection may be similar between fields the research background may produce some variability in the methods of training.

Table \ref{tab:percents} shows that responses from researchers who worked with several different scales of data were obtained. The research scale presents an interesting context for understanding the application of training in visual inspection. While key components of visual inspection may be similar, the application may be significantly different depending on whether an individual works with 1 object or 10000.

The aim of Questions 4 and 5 was to obtain responses from individuals with a high likelihood of requiring skills in visual inspection, while also providing some contextual information on their responses to the questions about training. The responses that have been collected suggest that this aim has been achieved.

\section{Training in Visual Inspection}\label{training in visual inspection}

Survey respondents were provided with a set of 6 questions that related to their experiences with training in visual inspection. These questions were designed to identify the prevalence and frequency of formal and informal training activities as well as the perceived relevance of the training to an individual's field of research. In addition, respondents were also asked to what extent providing formal or informal training was a part of their work as an astronomer.

As discussed in Section \ref{area of research}, responses in this section have been limited to the 70 respondents whose primary research area is observational.

\subsection{Formal and Informal Training}
\label{formal training}
Table \ref{tab:Formal} shows that a majority of respondents (79\%) indicated that they had not received formal training in visual inspection. Separating by career stage shows that 27\% of PGSs, 29\% of ECRs, 13\% of MCRs and 19\% of SCRs identified as receiving formal training.

\begin{table}[bt!]
    \centering
        \caption{\textit{Question 6: Have you ever received formal training in visual inspection of Astronomical images?} Responses have been grouped by career stage. This table has been limited to the 70 respondents from Observational Astronomy.}
        \label{tab:Formal}
    \begin{tabular}{|l|c|c|c|c|c|}
    \hline
        ~ & PGS & ECR & MCR & SCR &  Total \\ \hline
        No & 6 & 10 & 12 & 19 & 47 \\ \hline
        Not sure & 2 & 2 & 2 & 2 & 8 \\ \hline
        Yes & 3 & 5 & 2 & 5 & 15 \\ \hline
         Total & 11 & 17 & 16 & 26 & 70 \\ \hline
    \end{tabular}

\end{table}

Table \ref{tab:informal} shows that 60\% of respondents indicated that they had received informal training in visual inspection. Separating by career stage, 73\% of PGSs, 65\% of ECRs, 56\% of MCRs and 54\% of SCRs identified receiving informal training. 

The apparent decrease in engagement in both formal and informal training for later career stages could be due to several factors that are discussed in Section \ref{Engagement in Training in Visual Inspection}.

\begin{table}[bt!]
    \centering
    \caption{\textit{Question 7: Have you ever received informal training in visual inspection of Astronomical images?} Responses have been grouped by career stage. This table has been limited to the 70 respondents from Observational Astronomy.}
    \label{tab:informal}
    \begin{tabular}{|l|c|c|c|c|c|}
    \hline
        ~ & PGS & ECR & MCR & SCR &  Total \\ \hline
        No & 2 & 3 & 5 & 11 & 21 \\ \hline
        Not sure & 1 & 3 & 2 & 1 & 7 \\ \hline
        Yes & 8 & 11 & 9 & 14 & 42 \\ \hline
         Total & 11 & 17 & 16 & 26 & 70 \\ \hline
    \end{tabular}

\end{table}

A comparison of Tables \ref{tab:Formal} and \ref{tab:informal} demonstrates  that most of the training experiences of the survey respondents occurred in an informal manner. Additionally, most of the individuals who had received formal training also indicated that they had received informal training.

\begin{figure*}[bt!]

    \centering
    \includegraphics[clip,width = 0.48\linewidth]{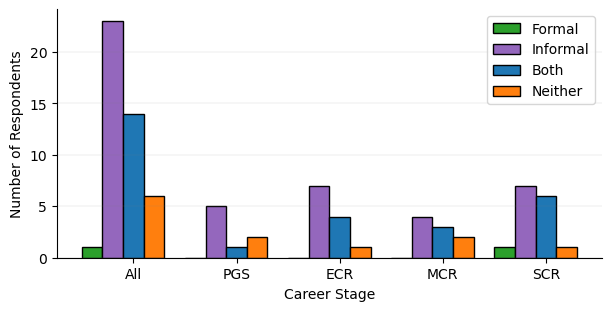}
    \includegraphics[clip,width = 0.48\linewidth]{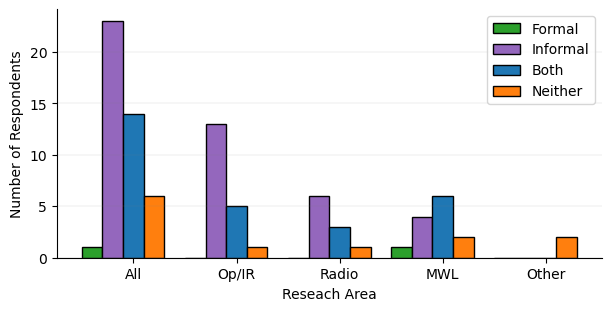}
    \caption{\textit{Question 6: Have you ever received formal training in visual inspection of Astronomical images?} and \textit{ Question 7: Have you ever received informal training in visual inspection of Astronomical images?} (left Panel) Responses sorted by career stage. (Right Panel) Responses sorted by research area where Optical/Infrared and Multi-Wavelength have been abbreviated to Op/IR and MWL respectively. This Figure has been limited to the 70 respondents from Observational Astronomy. In this figure we take the pessimistic stance that respondents who indicated that they were unsure whether they had received training are assumed to have not received training.}
    \label{circles}
\end{figure*}
Table \ref{tab:formalvinformal} also shows that 19 astronomers (27\% of respondents) indicated that they could not identify having received any training in visual inspection. An additional 10\% were unsure if they had ever received training in visual inspection. This suggests that as many as 37\% may not have been trained in visual inspection, despite self identifying as undertaking research activities classified as observational astronomy. While it is plausible that some individuals do not make use of visual inspection it seems unlikely that this could account for a quarter of observational astronomers, which suggests that a more complex relationship for skill acquisition is occurring. In addition, the responses gathered indicate that individuals at more senior career stages (MCR and SCR) were more likely to have not received training in visual inspection.
\begin{table}
    \caption{\textit{Question 6: Have you ever received formal training in visual inspection of Astronomical images?} And \textit{Question 7: Have you ever received informal training in visual inspection of Astronomical images?} Responses to Question 6 are reported in rows and responses to Question 7 are reported in columns. Together the table reports the combination of training respondents received.  This table has been limited to the 70 respondents from Observational Astronomy.}
    \centering
    \begin{tabular}{|l|l|c|c|c|c|}
    \hline
        ~&~&\multicolumn{3}{c|}{Informal Training}&~ \\ \hline
        ~&    ~& No & Not sure & Yes & Total \\ \hline
        ~ &No & 19 & 6 & 22 & 47 \\ \cline{2-6}
                            &Not sure & 0 & 1 & 7 & 8 \\ \cline{2-6}
        \multirow{-3.4}*{\rotatebox{90}{\begin{tabular}{c} Formal \\  Training\end{tabular}}}                    &Yes & 2 & 0 & 13 & 15 \\ \hline
                           ~ &Total & 21 & 7 & 42 & 70 \\ \hline
    \end{tabular}

    \label{tab:formalvinformal}
    
\end{table}

In Figure \ref{circles}, we take a pessimistic approach and assume that any individual who is not sure whether they had received training in fact did not receive training. 
The separation by research area (right panel) show that while Optical/Infrared astronomers accounted for only 53\% of respondents, they also accounted for 69\% of the astronomers who indicated that they had not received training. Across both career stages and research areas, informal training is the predominant method through which astronomers are trained.

\subsection{Recency of Training}
\label{frequency of training}
From Table \ref{tab:Timesformal}, 60\% of respondents who had received formal training indicated that their most recent training experience occurred at a time more than 5 years prior to being surveyed. Similarly from Table \ref{tab:Timesinformal}, 50\% of respondents who received informal training reported that their most recent training experience occurred at a time more than 5 years prior to being surveyed. In contrast, fewer than 15\% of the most recent formal training activities and 20\% of the most recent informal training activities had occurred in the last 12 months. Further, by excluding the responses of PGSs, who might be expected to receive frequent training through their postgraduate studies, only 3 individuals indicated receiving formal training in the last 5 years and only 4 individuals indicated receiving informal training in the last 12 months.

\begin{table}[bt!]
    \caption{\textit{Question 8: When did you last receive formal training in visual inspection?} This table has been limited to the 15 respondents from Observational Astronomy who indicated they had received formal training in visual inspection in Question 6.}
    \centering
    \begin{tabular}{|l|c|c|c|c|c|}
    \hline
        ~ & PGS & ECR & MCR & SCR & Total \\ \hline
        $<$12 months & 1 & 0 & 0 & 1 & 2 \\ \hline
        $<$5 years & 2 & 0 & 1 & 1 & 4 \\ \hline
        $>$5 years & 0 & 4 & 2 & 3 & 9 \\ \hline
        Total & 3 & 4 & 3 & 5 & 15 \\ \hline
    \end{tabular}

        \label{tab:Timesformal}
\end{table}

\begin{table}[bt!]
    \caption{\textit{Question 9: When did you last receive informal training in visual inspection?} This table has been limited to the 42 respondents from Observational Astronomy who indicated they had received informal training in visual inspection in Question 7.}
    \centering
    \begin{tabular}{|l|c|c|c|c|c|}
    \hline
        ~ & PGS & ECR & MCR & SCR & Total \\ \hline
        $<$12 months & 4 & 3 & 0 & 1 & 8 \\ \hline
        $<$5 years & 4 & 4 & 2 & 1 & 11 \\ \hline
        $>$5 years & 0 & 3 & 7 & 11 & 21 \\ \hline
        Not sure & 0 & 1 & 0 & 1 & 2 \\ \hline
        Total & 8 & 11 & 9 & 14 & 42 \\ \hline
    \end{tabular}

        \label{tab:Timesinformal}
\end{table}

\subsection{Relevance of Training Activities}

\label{training relevance}
From Table \ref{tab:relevance}, a slight majority (52\%) of individuals indicated their most recent training experience had high relevance to their current work. A further 32\% indicated moderate relevance, 14\% indicated very high relevance and 2\% indicated low relevance. The only individual who indicated low relevance identified transferring between two fields of observational astronomy, which could explain why their training no longer relates to their work. Importantly, despite being given the option, no respondent indicated the training they received had ``no relevance'' to their current research.

Individuals who indicated they received both formal and informal training tended to respond that their training had higher relevance. Out of the 13 respondents who received both types of training, 25\% indicated very high relevance, 69\% indicated high relevance and 6\% indicated moderate relevance. 

\begin{figure*}

    \centering
    \includegraphics[trim = 1.6cm 0.3cm 1.8cm 0.4cm,clip, width = 0.49\linewidth]{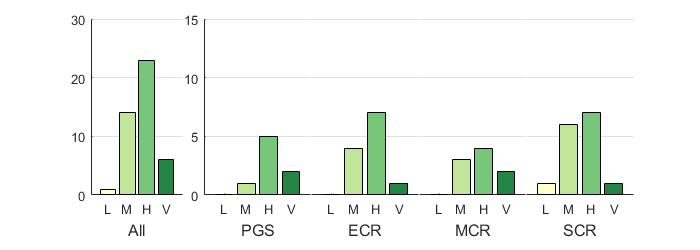}
    \includegraphics[trim = 1.6cm 0.3cm 1.8cm 0.4cm,clip, width = 0.49\linewidth]{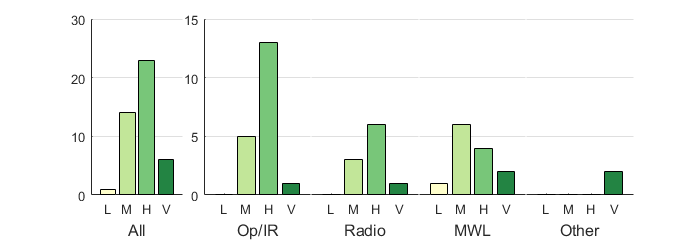}
    \caption{\textit{Question 10: How relevant to your current work is the formal or informal training you have received?} (left Panel) Responses sorted by career stage. (Right Panel) Responses sorted by research area where Optical/Infrared and Multi-Wavelength have been abbreviated to Op/IR and MWL respectively. This figure has been limited to the 44 respondents from Observational Astronomy who received either formal or informal training. The response options are represented in each graph from left to right: (L) Low relevance, (M) Moderate relevance, (H) High relevance and (V) Very high relevance.}\label{training relevance bars}
\end{figure*}

In Figure \ref{training relevance bars} responses, regarding relevance are separated by both career stage and research area. In the case of respondents from both Optical/Infrared and Radio Astronomy research areas, the majority of respondents indicated high relevance(68\% and 60\% respectively) for the training they received. In contrast, 46\% of respondents in multi-wavelength research indicated moderate relevance and only 31\% indicated high relevance. These results present a platform for the discussion in section \ref{Methods of Training in Visual Inspection and their value}.

\begin{table}
    \centering
    \caption{\textit{Question 10: How relevant to your current work is the formal or informal training you have received?} Responses have been shortened to be presented in a table. The full version of the options presented to survey participants were: Very high (I would not be able to complete my current work without the training I received); High relevance (Elements of my current work rely on the training I received); Moderate relevance (My training assists my current work but it is not essential); and Low relevance (I have taken elements from the training but, overall it is not necessary for my current work). This table has been limited to the 44 respondents from Observational Astronomy who indicated they had received some form of training in visual inspection in either Question 6 or 7.}
    \label{tab:relevance}
    \begin{tabular}{|l|c|c|c|c|c|}
    \hline
        ~ & PGS & ECR & MCR & SCR &  Total \\ \hline
        Very high relevance & 2 & 1 & 2 & 1 & 6 \\ \hline
        High relevance & 5 & 7 & 4 & 7 & 23 \\ \hline
        Moderate relevance & 1 & 4 & 3 & 6 & 14 \\ \hline
        Low relevance & 0 & 0 & 0 & 1 & 1 \\ \hline
        Total & 8 & 12 & 9 & 15& 44 \\ \hline
    \end{tabular}
\end{table}

\subsection{The Provision of Training in Visual Inspection}
\label{the provision of training in visual inspection}
The final 3 questions all relate to the providers of training. For Question 11, respondents who indicated that they had received informal training were asked to identify the role of the individual who provided their training. All survey respondents were asked to identify whether providing formal or informal training in visual inspection was a component of their work as a researcher in Questions 12 and 13. 

In Table \ref{tab:who are trainers}, respondents indicated that most informal training was performed by either a member of their PhD supervisory team (35\%) or an academic who works in the same field as the trainee (31\%). The only other category of trainer that was identified by more than 10\% of respondents was ``other academics'' (17\%). Due to respondents identifying only their most recent training experience, PhD supervisors may be under-reported as trainers in these responses: later career researchers (MCR and SCR) might be expected to have undergone training experiences more recently than during their postgraduate training.

\begin{table}[bt!]
    \caption{\textit{Question 11: Consider your most recent experience receiving informal training in visual inspection of astronomical images.  From the following list, please select the most accurate description of the role of the person who provided that informal training.}  This table has been limited to the 42 respondents from Observational Astronomy that indicated they had received informal training in Question 6. }
    \centering
    \begin{tabular}{|p{0.15\textwidth}|c|c|c|c|c|}
    \hline
        ~ & PGS & ECR & MCR & SCR & Total \\ \hline
        A member of my PhD supervisory team & 6 & 3 & 3 & 3 & 15 \\ \hline
        A PhD student peer & 0 & 0 & 1 & 1 & 2 \\ \hline
        A postdoctoral research fellow & 1 & 0 & 1 & 0 & 2 \\ \hline
        An academic peer or colleague & 0 & 1 & 1 & 5 & 7 \\ \hline
        An academic researcher from my research area & 1 & 4 & 3 & 5 & 13 \\ \hline
        I cannot recall & 0 &3 & 0 & 0 & 3 \\ \hline
         Total & 8 & 11 & 9 & 14 & 42 \\ \hline
    \end{tabular}

    \label{tab:who are trainers}
\end{table}

\begin{figure*}[bt!]

    \centering
    \includegraphics[ width = 0.98\linewidth]{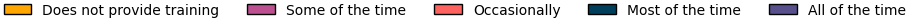}
    \includegraphics[trim = 1.6cm 0.3cm 1.8cm 0.4cm,clip, width = 0.49\linewidth]{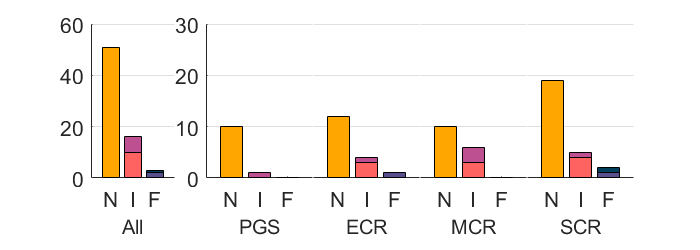}
    \includegraphics[trim = 1.6cm 0.3cm 1.8cm 0.4cm,clip, width = 0.49\linewidth]{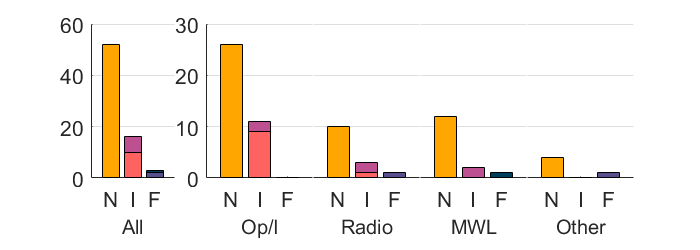}
    \includegraphics[trim = 1.6cm 0.3cm 1.8cm 0.4cm,clip, width = 0.49\linewidth]{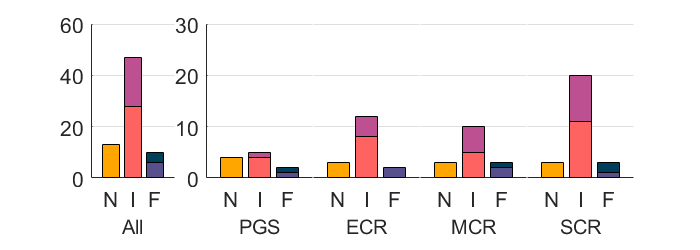}
    \includegraphics[trim = 1.6cm 0.3cm 1.8cm 0.4cm,clip, width = 0.49\linewidth]{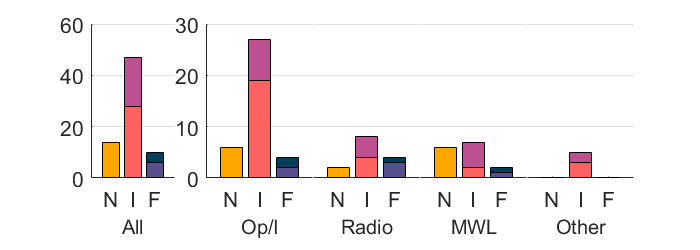}

    \caption{\textbf{[Top] }\textit{Question 12: To what extent do your current (or most recently completed) duties require you to deliver formal training in visual inspection activities?} \& \textbf{[Bottom] }\textit{Question 13: To what extent do your current (or most recently completed) duties require you to deliver informal training in visual inspection activities?}(left Panels) Responses sorted by career stage. (Right Panels) Responses sorted by research area where Optical/Infrared and Multi-Wavelength have been abbreviated to Op/IR and MWL respectively. This table has been limited to the 70 respondents from Observational Astronomy. Responses are separated into three categories: (N) Non-trainers, those who indicated that they did not provide training, (I) Infrequent trainers, which includes respondents who indicated they provided training ``some of the time''(top of stack) and ``occasionally''(bottom of stack) and (F) Frequent trainers, which includes respondents who said they provided training ``most of the time''(top of stack) and ``all the time''(bottom of stack).}
    \label{Supply formal training multi}
\end{figure*}

In Table \ref{tab:supply formal training all} and Figure \ref{Supply formal training multi} it can be seen that a small number of respondents do perform formal training. Out of the 19 respondents who answered that they provide formal training, only 3 indicated that they did so frequently. The majority of formal trainers (69\%) were part of the Optical/Infrared cohort, those 19 respondents accounted for 30\% of that cohort. MCRs reported the highest percentage of delivering formal training, 38\%, followed by ECRs (30\%), SCRs (27\%) and finally PGSs (9\%).
\begin{table}[ht!]
    \centering
        \caption{\textit{Question 12: To what extent do your current (or most recently completed) duties  require you to deliver formal training in visual inspection activities?} This table has been limited to the 70 respondents from Observational Astronomy. }
    \label{tab:supply formal training all}
    \begin{tabular}{|p{0.06\textwidth}|p{0.09\textwidth}|c|c|c|c|c|}
    \hline
        ~&~ & PGS & ECR & MCR & SCR & Total \\ \hline
        ~&All of the time  & 0 & 0 & 0  & 1  & 1  \\ \cline{2-7}
        \multirow{-2.5}{*}{\rotatebox{45}{Frequent}}  &Most of the time  &  0 & 1 & 0  & 1  & 2  \\ \hline
        ~&Some of the time  & 1  & 1 & 3  & 1  & 6  \\ \cline{2-7}
        \multirow{-2.5}{*}{\rotatebox{45}{Infrequent}} & Occasionally  \textcolor{white}{test}  &  0 & 3 & 3  & 4  & 10  \\ \hline
        Not at all&~  & 10  & 12 & 10  & 19  & 51  \\ \hline
    \end{tabular}
\end{table}

\begin{table}[ht!]
    \centering
            \caption{\textit{Question 13: To what extent do your current (or most recently completed) duties require you to deliver informal training in visual inspection activities?} This table has been limited to the 70 respondents from Observational Astronomy.}
    \label{tab:supply informal training all}
    \begin{tabular}{|p{0.06\textwidth}|p{0.09\textwidth}|c|c|c|c|c|}
    \hline
        ~& ~ & PGS & ECR & MCR & SCR & Total \\ \hline
       ~  &All of the time  & 1  & 0 & 1  & 2  & 4  \\ \cline{2-7}
        \multirow{-2.5}{*}{\rotatebox{45}{Frequent}}&Most of the time  & 1  & 2 & 2  & 1  & 6  \\ \hline
        ~ &Some of the time  & 1  & 4 & 5  & 9  & 19  \\ \cline{2-7}
        \multirow{-2.5}{*}{\rotatebox{45}{Infrequent}} &Occasionally \textcolor{white}{test} & 4  & 8 & 5  & 11  & 28  \\ \hline
        Not at all & ~  & 4  & 3 & 3  & 3  & 13  \\ \hline
    \end{tabular}
\end{table}
In comparison, Table \ref{tab:supply informal training all} indicates that most respondents from observational astronomy identified that they provided informal training in visual inspection as a component of their work as a researcher. 14\% identified themselves as a frequent provider of informal training and 67\% identified themselves as an infrequent provider of informal training. 

Interestingly, the responses in Table \ref{tab:supply informal training all} and Figure \ref{Supply formal training multi} do not completely align with those in Table \ref{tab:who are trainers}. In Table \ref{tab:who are trainers} respondents indicated the people who predominantly provide informal training are PhD supervisors, while only 2 individuals indicated being trained by a Postgraduate student. From Table \ref{tab:supply informal training all}, however, 64\% of PGS suggested that providing informal training in visual inspection was a component of their work as a researcher. In addition, more individuals indicated that they provided formal training than received any kind of training

By considering research area cohorts, Figure \ref{Supply formal training multi} also shows that the PGSs who indicated that they provided informal training were not limited to one research area. Additionally, Figure \ref{Supply formal training multi} also shows that is that providing informal training was less common in the multi-wavelength cohort compared to the others.

\begin{table}[bt!]
    \caption{\textit{Question 12: To what extent do your current (or most recently completed) duties require you to deliver formal training in visual inspection activities?} This table has been limited to the 26 respondents from Observational Astronomy who indicated they had not received training in visual inspection in either Question 6 or 7.}
    \centering
    \begin{tabular}{|l|c|c|c|c|c|}
    \hline
        ~ & PGS & ECR & MCR & SCR & Total \\ \hline
        Most of the time & 0 & 0 & 0 & 1 & 1 \\ \hline
        Some of the time  & 0 & 0 & 3 & 0 & 3 \\ \hline
        Occasionally & 0 & 0 & 1 & 2 & 3 \\ \hline
        Not at all & 3 & 5 & 3 & 8 & 19 \\ \hline
        Total & 3 & 5 & 7 & 11 & 26 \\ \hline
    \end{tabular}

\label{tab:untrainedformaltrainers}
\end{table}

\begin{table}[bt!]
\caption{\textit{ Question 13: To what extent do your current (or most recently completed) duties require you to deliver informal training in visual inspection activities?} This table has been limited to the 26 respondents from Observational Astronomy who indicated they had not received training in visual inspection in either Question 6 or 7.}
\label{tab:untrainedinformaltrainers}
    \centering
    \begin{tabular}{|l|c|c|c|c|c|}
    \hline
        ~ & PGS & ECR & MCR & SCR & Total \\ \hline
        All of the time & 0 & 0 & 0 & 1 & 1 \\ \hline
        Most of the time & 0 & 1 & 1 & 0 & 2 \\ \hline
        Some of the time  & 0 & 2 & 4 & 2 & 8 \\ \hline
        Occasionally & 1 & 1 & 1 & 6 & 9 \\ \hline
        Not at all & 2 & 1 & 1 & 2 & 6 \\ \hline
        Total & 3 & 5 & 7 & 11 & 26 \\ \hline
    \end{tabular}
\end{table}

The most important subset of responses are those 19 (27\% of the observational astronomy cohort) identified in Table \ref{tab:formalvinformal} who definitely indicated that they had received no training in visual inspection, a task that we propose is a critical component of visual discovery which in turn is critical for observational astronomy. Limiting responses to Questions 12 and 13, as can be seen in Tables \ref{tab:untrainedformaltrainers} and \ref{tab:untrainedinformaltrainers}, 31\% respondents indicated providing some formal training and 79\% indicated that they provided some informal training. Based on their responses, these individuals are training others in a task that they themselves appear to have never been taught to do.

\section{Interview on Individual Experiences of Training}
\label{interviews on iddividual experience of training}
 While we anticipated that a survey would give a good macroscopic understanding of these focus areas, three pilot interviews were also undertaken to further understand the individual experience of training.

The final four interview questions were selected due to their ability to supplement key survey results. The interview questions were:
\begin{enumerate}
    \item Can you please provide one specific example as to how you have been trained in the visual inspection of images? Are there any other examples you would like to share?
    \item Are you happy with the types of training you have received? Why/why not?
    \item At this stage of your career, would a retraining program in visual inspection of images be of value to you? Why/why not?
    \item Consider your experiences more broadly than visual inspection of images. What is the best training experience you have had? What features made it the best training experience?
\end{enumerate}

These interview questions also provide a framework against which the reader can analyze their own experiences with training in visual inspection in astronomy. We encourage the reader to consider their personal responses to these questions before reading those presented in this paper.

\subsection{Rationale}
The interviews help to develop a dialogue through which specific training experiences can be discussed. This was done in the following two key ways.

Firstly, question 1 helps us to attain a more complete understanding of the types of training astronomers received, and how those training experiences  relate to the concepts of formal and informal training.

Secondly, questions 2, 3 and 4 help provide a more in depth understanding of the value of training. In survey question 10, respondents indicated that the training they had received had a high relevance. Interview question 2 provides insight into the satisfaction associated with training. Question 3 helps identify whether astronomers consider visual inspection to be a skill that requires maintenance and retraining. Question 4 provides insight into the key features that are associated with good training, which provides an important counterpoint to question 10. 

\subsection{Methods}
The pilot interviews were conducted using a structured interview model. No additional questions were asked beyond those listed at the start of Section \ref{interviews on iddividual experience of training}. However, follow-up prompts such as ``why/why not?'' were used to gain richer responses. Notes were taken during the interviews, however, the audio of the interviews was not recorded as per the approved ethics protocol. Interviews were conducted using the Zoom platform. Two researchers were present during each interview and separately noted responses that were discussed post interview. In the analysis below, both a thematic review of responses, overarching themes between responses, as well as areas of significance in individual responses, are provided.

An option to participate in pilot interviews was presented to respondents during the first two rounds of advertising  to the ASA in June and July 2021. Out of the initial 20 survey respondents, 3 consented to be interviewed.

An internal review was performed that took into consideration the number of interview responses, the responses themselves and the logistics of performing further interviews at the height of the Covid-19 pandemic. Consequently, a decision was made to discontinue the interview component on the 4th of August 2021. However, the 3 interviews collected provide a valuable counterpoint to the elements discussed in Section \ref{Knowledge}  and Sections \ref{formal training} and \ref{the provision of training in visual inspection}.

\subsection{Participant Profiles}
\label{participant profiles}
The interviews captured responses from individuals with a range of career stages and research backgrounds. Below is a brief description of each interview subject that reflects their career stage at the time of their interview.

\begin{itemize}
    \item Subject A was a student studying for their PhD (PGS). They predominantly work in the area of observational astronomy.
    \item Subject B was an active researcher (MCR). In addition to the training they have received throughout their career, they have also engaged in the provision of training.
    \item Subject C was working in a field outside of observational astronomy (SCR). They had worked in astronomy for the majority of their research career.
\end{itemize}
\subsection{Question 1}
\label{question 1}

All three interview participants indicated that their training in visual inspection began with or included a significant component of informal mentor-mentee training. Most notably subject A described a ``Senior Researcher'' and subject B described an ``Expert''. All three interviewees also identified a large amount of training throughout their experience as a researcher consisted of working with data to better understand what they were looking for in data. Subject B concisely defined this with the point ``I just figured out what I was looking for'' in relation to visual inspection tasks. Both subjects B and C acknowledged receiving training from a PhD student peer. 

Subject A described a workshop organised by their PhD supervisor as a principal factor in developing their research skills. This was the clearest example of formal training that was provided by any of the respondents. Subject A said that this training exercise was repeated for all group members each time a new member joined. They stated that because of this, training activities had reduced value over multiple repetitions.

Subject B described an experience from their postgraduate studies in which they performed quality assurance for developing an automated system. Their example included an element where an expert supplied training but also included a large portion of training through repetition.

\subsection{Question 2} 
\label{question 2}

All three interviewees indicated some dissatisfaction with the training they had experienced. Subjects A and B described a lack of support, reporting that they had been ``Thrown in the deep end''. Subjects B and C discussed preferring work-integrated learning over being told how to perform a task. 

Subject A expanded on their perceived dissatisfaction with repetition in training that was identified in their earlier response. They again discussed the limited knowledge improvement as training was repeated with a specific emphasis on the lack of value in re-watching recorded sections.

Subject B's responses provided an important perspective because of their experience as an educator. They emphasised the challenges of being an educator for new researchers given the lack of preparation for teaching. They discussed difficulties in training others because astronomy has dense topics that ``Cannot be broken up into bits''. Similarly, they struggled with empathising with the beginner's perspective, to ``put yourself in a beginner's shoes''.

Subject C discussed challenges with induction and onboarding processes. They identified several causes for this, the most prevalent of which was the language used which they described as ``Jargon'' and differences in prior knowledge between trainers and trainees. 

\subsection{Question 3}
\label{question 3}

All three interviewees expressed a lack of interest in retraining programs in general. The overarching reason for this was a perceived expertise in the types of visual inspection used in their field of research. Both Subjects A and B indicated some interest in retraining if it correlated with training in a novel visual inspection task. 
 
Subject B cited the decade of experience in their field of research as a key reason why retraining would have little to no value to them. Subject B referred to the quality assurance task that they had trained in, noting that it was now an outdated practice. As such the training they had received was, and hence retraining in that task would be, redundant. 

Subject C noted that at their current career stage learning new techniques for visual inspection would have little value. They expressed that their interest had moved from visual inspection per se towards the systems and tools used for visual inspection. Additionally, they identified that they were highly proficient in the systems that they used. Due to these factors, they expressed no interest in training in new systems or visual inspection techniques as these would add little value to their work.
\subsection{Question 4}
\label{question 4}

The responses to this question were highly individualised. The only common elements were an interest in written elements (subjects A and B) and collaboration (Subjects A and B).

Subject A noted the importance of streamlining presentations through the use of supplementary written documents possibly because they allow the learner to train at their own pace.

Subject B discussed the importance of both explanations and a demonstration of a given task specifically citing a need for ``hands-on demonstrations with lots of feedback''. They discussed the value of collaboration noting the opportunity for peer feedback.

Subject C discussed the importance of content being ``Pitched at the correct level''. They further discussed the importance of the individualization of the training that was provided.

\section{Discussion}
\label{summary and discussion}
In Sections \ref{training in visual inspection} and \ref{interviews on iddividual experience of training}  we presented the results gathered in a survey, supported by pilot interviews, that was completed by members of the Astronomical Society of Australia and the International Astronomical Union between June 2021 and May 2022. 

The objective of this work was to take a sample of the observational astronomical community in order to investigate aspects of their personal experiences relating to training in visual inspection of images. We discuss our findings from the survey, limiting our sample to the 70 observational astronomers who participated, and the three pilot interviews with regards to these four themes.

\subsection{Methods of Training in Visual Inspection and their value}
\label{Methods of Training in Visual Inspection and their value}

The results from this survey show that the majority of the observational astronomy cohort, 63\%, indicated they had received some form of training in visual inspection. A further 10\% were not sure they had received training. This training was generally informal and delivered by either a PhD supervisor or an academic in the same field of research as the trainee. Additionally, we found that respondents indicated their most recent training occurred a significant time in the past, 50\% of formally trained respondents and 60\% of informally trained respondents indicated that their training occurred more than 5 years prior to being surveyed. 

As might be expected, training appears to be predominantly limited to the early parts of a researcher's career. The postgraduate research phase offers direct access to established researchers which allows for opportunities for informal mentor-mentee training. From the responses gathered it is clear that the PhD supervisor plays a critical role in the provision of training, with 36\% indicating that their most recent training experience was delivered by their PhD supervisor. This could even be an underestimate given respondents from later career stages may have received training post PhD.

While training is certainly most critical in the early stages of an individual's career it is not necessarily the only stage where it is applicable. For example, some visual inspection task that is not frequently repeated. In such instances skill loss between repetitions may not be obvious to the individual performing the task. Skill loss as a result of lack of use \citep[see][]{ArthurJr98} or as a consequence of automation \citep[see][]{Bainbridge83} is not new. Training programs that emphasise skill maintenance have been designed \citep[see][]{Kluge14} and tested \citep[see][]{Frank18}. However, it is unclear if there are training programs that cater for this kind of professional development in observational astronomy.

It appears that best practices, or even the most common practices, for training in visual inspection have not been documented or shared widely amongst the observational community.   The benefit of documenting such skill-based training approaches is the opportunity to investigate, understand and make comparisons between approaches.  For example, it is not currently possible to examine and assess the training methods utilised by individual PhD supervisors and those of more structured activities, such as the project-specific training described by interview Subject A (see Section \ref{interviews on iddividual experience of training}).

Those who provide training might be better, worse or different depending on their experience, knowledge and ability to articulate concepts. Likewise, astronomers may learn more effectively, or be more satisfied with their learning, depending on the teaching approach that is used. As with publishing novel approaches to scientific problems, there is value in also presenting novel approaches to teaching visual inspection. Where literature can be found regarding training in astronomy, it tends to focus on: (1) the acquisition of knowledge (relevant for the interpretation phase of visual discovery); and (2) the educational needs of primary and secondary students rather than professionals. Interview subject B summarised this challenge as a lack of preparation for teaching or more concisely, a need for training in how to train others. The first step to this is uncovering how we currently train astronomers in visual inspection.

\subsection{Engagement in Training in Visual Inspection}
\label{Engagement in Training in Visual Inspection}

If the results collected in Section \ref{formal training} are reflective of the wider astronomy community then they are somewhat concerning: almost 30\% of respondents indicated that they had not received training in visual inspection. It is important to recognise that untrained is not the same as unskilled. We in no way suggest that 30\% of astronomers cannot perform visual inspection. It is unclear, however, how or when they developed those skills.

This result emphasises the different perspectives regarding skills and knowledge in astronomy. It is expected that an astronomer would keep abreast of developments within their field of study, maintaining up-to-date knowledge that could assist in making, or perhaps more importantly, interpreting or explaining a discovery. We could not find any clear published evidence that a reference guide of methods of visual inspection exists within the context of astronomy, however, \citet{Kok15} and \citet{Richter20} show that within the context of medical imaging, there is a hierarchy of inspection strategies, from systematic to non-systematic viewing.

Within this work, visual inspection is presented as the act of examining an image to identify areas or objects of interest. While visual inspection is a common part of observational astronomy, it does not appear to be one that is commonly discussed. This simple definition was presented to assist survey respondents in gauging their responses, however, the definition can be extended using the model of biologically primary and secondary knowledge \citep{Geary08}. In this model, biologically primary knowledge is knowledge which we are innately predisposed to acquire while secondary knowledge requires time and cognitive resources to develop. Applying this model to visual inspection of images, we identify: (1) pattern recognition as biologically primary; and (2) understanding of the contextual background of astronomy as secondary knowledge. 

Further, the work of \citet{Tricot13} suggests that biologically primary abilities are acquired easily through use and automatically without the necessity of instruction. This is consistent with the responses from the interview participants, who suggested that a large portion of training in visual inspection occurred through ``figuring out'' rather than guided or structured training. This may also explain why 30\% of the observational astronomer cohort indicated they had not received training, as it would be consistent with a component of self directed skill acquisition. 

However, this interpretation presents an additional challenge for the future: if skills in visual inspection are acquired through use, how will future astronomers -- especially postgraduate and early career researchers -- gain this ability in an era when automation is implemented to a greater extent (see Section \ref{automation})? And in turn, when these postgraduate students and early career researchers become mid-career researchers and senior career researchers, how will they pass on these skills to their own students?

\subsection{The Value of Training}
From the responses in Figure \ref{training relevance bars} it is clear that respondents who engage in training regard it as being highly relevant to their work. This is a positive result that suggests that the outcome of training is the development of useful skills. In contrast, our interview participants indicated a general dissatisfaction towards the training they received. Based on their descriptions, this dissatisfaction was related to the method of the training rather than the content of the training. If we speculate that within the wider community, there is a general trend of high relevance but low satisfaction in regards to training in visual inspection it would suggest a larger problem with the method in which we train visual inspection.

Relevance and satisfaction are only two components of the value of training.  Ultimately, what is more important to understand is the effect training has on an individual's ability to perform. 
Understanding how factors such as training mode (formal versus informal) or the recency of training actually affect performance is beyond the scope of this present work. Further investigation with a specific focus on these aspects of skill acquisition for visual inspection is warranted. 

In addition, future investigations must address how the value of the training provided today can be judged within the context of how visual inspection will likely be used in observational astronomy in the highly-automated future.

\subsection{Trainers without Precedent}
\label{those who teach}

Perhaps the most unexpected results from this survey were found in Section \ref{the provision of training in visual inspection}. A comparison of Tables \ref{tab:informal} and \ref{tab:supply informal training all} presents a striking result: more respondents provided informal training in visual inspection than received informal training in visual inspection. Likewise, a comparison of Tables \ref{tab:Formal} and \ref{tab:supply formal training all} presents a similar relationship between providing and receiving formal training in visual inspection. This suggests a mismatch between respondents' perceived engagement in training as a trainer compared to a trainee.

It should be noted that respondents were not provided with a definition for the response options. For instance, a respondent may consider 1 hour of training per week to be ``occasionally'' whereas another may consider 1 hour of training per week to be ``some of the time''. As such, the responses in Tables \ref{tab:supply formal training all} and \ref{tab:supply informal training all}  are considered in terms of frequent trainers (all of the time and most of the time), infrequent trainers (some of the time and occasionally) and non-trainers (not at all).

One possible explanation for the mismatch in the perceived provision and receipt of training is that training is more likely to be remembered by the provider.
More specifically, while a trainer might remember all the training they provided, a trainee may only recall the highly valuable training they received. From Tables \ref{tab:Timesformal} and \ref{tab:Timesinformal} and Figure \ref{training relevance bars} it was suggested that training in visual inspection is highly relevant but occurs in the early stages of an observational astronomer's career. Alternatively, it could also be the training that is received in the later stages of a career are less important and therefore less likely to be remembered. 

Additionally, when discussing trainers who have never been trained it is possible that they have attained ``expertise without precedent''. In every field there must be an individual or individuals who were the first to perform that work. In these instances, there was no available method of training, and so individuals self-train. One avenue through which these skills could be developed is via biologically primary and secondary means (see Section \ref{Methods of Training in Visual Inspection and their value}). However, in such instances, how can the effectiveness of methods learnt be assessed?

Here again, the paucity of literature on how training in visual inspection is delivered in astronomy prevents us from disentangling these two explanations. Without knowing the methods through which researchers at different stages in their careers are trained, and if these methods of training differ, it is not possible to determine whether the challenge in training is a value problem, an accessibility problem or a motivation problem.

\section{Considerations for the Future}
\label{final thoughts}

In this work, we investigated the training in visual inspection through a survey of 70 observational astronomers. We found that the majority of respondents (60\%) received informal training. However, responses also indicated that 27\% had not received formal or informal training in visual inspection. we found that the majority of those who received informal training in visual inspection considered that training to be highly relevant to their work (14\% very high relevance, 52\% high relevance). Additionally, the majority of the training was identified as being provided by either a PhD supervisor (36\%) or an academic researcher within their field (30\%). Lastly, we found that survey respondents provided significantly more informal training (81\%) than formal training (27\%).

Work investigating how medical diagnosticians perform and train visual inspection (e.g. \citet{Manning06}, \citet{Kok15}, \citet{vanderGijp16}, and \citet{Richter20}) is abundant.  There is certainly a wide scope of potential avenues in which future investigation on visual inspection in astronomy could take which build upon these previous works: (1) The distinction between systematic and non-systematic visual inspection as well as the lack of connection between inspection pattern and performance; (2) Massed practice or learning by doing and its relation to expertise and performance; and (3) The education of novices in systematic viewing and the trend towards non-systematic viewing in more senior observers. However further investigation into training in visual inspection is both difficult and time sensitive. We propose two critical challenges in this research space that need to be addressed in the near future.

Firstly, while those senior astronomers who have performed the most inspection during their careers are still within their field of research, what can their skills and abilities in visual inspection tell us about the best practices for developing skills in visual inspection?

Secondly, how can we implement training in visual inspection as automation increases and opportunities for training decrease?

The reality is that, in astronomy, the scale of data is changing. The rate of data collection, the size of collected data and the scale of projects are all increasing. While the technology used to collect data and make discoveries is always improving, it is undeniable that the scale of data is too great for astronomers to manage without external assistance. Automating elements of the discovery process is now an essential part of how research must be conducted in the future. However, the skills generations of astronomers have developed in visual inspection may be at risk: If we consider that more senior astronomers are likely to have performed visual inspection tasks more frequently it may be that the astronomers who have looked at the most astronomical images is in the late stages of their career. There is a finite amount of time before these experts leave the field, in which a decision has to be made on whether or not the astronomical profession values those skills and how we retain -- and train -- them for the future.

\section{Declarations}
\subsection{Ethical Considerations}
This research was approved by Swinburne University of Technology through the Swinburne University Human Research Ethics Committee (SUHREC) as per approval number 20236928-13912.
\subsection{Funding}
We acknowledge funding and support through the Australian Government Research Training Program Scholarship. During the period that this research activity was undertaken, Christopher Fluke was the SmartSat Cooperative Research Centre (CRC) Professorial Chair of space system real-time data fusion, integration and cognition. This work has been supported by the SmartSat CRC, whose activities are funded by the Australian Government’s CRC Program.
\subsection{Acknowledgements}
We would like to thank the editing team at the Astronomy Education Journal as well as the reviewers for taking the time to examine this work and provide feedback.

We acknowledge the Wurundjeri people of the Kulin Nation as the traditional owners of the land on which this research was conducted.


\bibliography{main.bib}

\end{document}